# A continuum model for the unfolding of von Willebrand Factor

Mansur Zhussupbekov[1], Rodrigo Méndez Rojano[1], Wei-Tao Wu[2], Mehrdad Massoudi[3], James F. Antaki[1]

[1] Department of Biomedical Engineering, Cornell University, Ithaca, NY, USA
[2] Department of Aerospace Science and Technology, Nanjing University of Science and Technology, Nanjing, China
[3] U. S. Department of Energy, National Energy Technology Laboratory (NETL), Pittsburgh, PA, USA

*Address for correspondence:*

James F. Antaki, PhD
Weill Hall 109
Meinig School of Biomedical Engineering
Cornell University
Ithaca, NY 14853
Phone: 607-255-0726
Email: antaki@cornell.edu

## Abstract

von Willebrand Factor is a mechano-sensitive protein circulating in blood that mediates platelet adhesion to subendothelial collagen and platelet aggregation at high shear rates. Its hemostatic function and thrombogenic effect, as well as susceptibility to enzymatic cleavage, are regulated by a conformational change from a collapsed globular state to a stretched state. Therefore, it is essential to account for the conformation of the vWF multimers when modeling vWF-mediated thrombosis or vWF degradation. We introduce a continuum model of vWF unfolding that is developed within the framework of our multi-constituent model of platelet-mediated thrombosis. The model considers two interconvertible vWF species corresponding to the collapsed and stretched conformational states. vWF unfolding takes place via two regimes: tumbling in simple shear and strong unfolding in flows with dominant extensional component. These two regimes were demonstrated in a Couette flow between parallel plates and an extensional flow in a cross-slot geometry. The vWF unfolding model was then verified in several microfluidic systems designed for inducing high-shear vWF-mediated thrombosis and screening for von Willebrand Disease. The model predicted high concentration of stretched vWF in key regions where occlusive thrombosis was observed experimentally. Strong unfolding caused by the extensional flow was limited to the center axis or middle plane of the channels, whereas vWF unfolding near the channel walls relied upon the shear tumbling mechanism. The continuum model of vWF unfolding presented in this work can be employed in numerical simulations of vWF-mediated thrombosis or vWF degradation in complex geometries. However, extending the model to 3-D arbitrary flows and turbulent flows will pose considerable challenges.

## Introduction

In recent years, the role of von Willebrand Factor (vWF), a multimeric glycoprotein circulating in blood, has been established as the primary mediator of high-shear thrombosis.[10,46,49,51] Numerous in vitro and in vivo tests have demonstrated the exclusive requirement of vWF for high-shear platelet aggregation.[50,52,69,71] Pathological and immunohistochemical examination of the thrombi found in explanted HeartMate II left ventricular assist devices (LVAD), for example, revealed that the inner rings of the laminated clots were composed primarily of fibrin and vWF.[7] Not only is vWF the culprit in the problem of thrombosis but also a victim in another: bleeding events in LVAD patients have been linked to the loss of large-molecular-weight vWF proteins.[39,47] vWF degradation in LVAD environment occurs due to mechanical demolition by supra-physiologic shear levels and excessive enzymatic cleavage by metalloprotease ADAMTS-13.[6,9,13]

In physiologic flow conditions, vWF multimers normally circulate in plasma in a compact globular configuration which prevents undesired adhesion since the binding sites remain hidden. Both vWF binding to collagen or platelets and its mechanoenzymatic cleavage are preceded by the unfolding of the vWF chain.[19,35,65,76] When the chain is extended, the tensile force exposes and activates the vWF-A1 domain that binds to collagen and GPIbα on the platelet membrane. In large vWF concatemers, this also exposes the A2 domain enabling site-specific cleavage by ADAMTS-13. These force-dependent activation and size regulation mechanisms are intended to localize and restrict the profound hemostatic potential of vWF to the site of injury. In pathologic conditions, however, this fine tuning is disrupted resulting in acute thrombogenesis or excessive vWF degradation leading to the acquired von Willebrand syndrome (aVWS).[29,48,70]

In hydrodynamic simulations of a single polymer in simple shear, a coiled vWF has been predicted to unfold at 5000 s$^{-1}$.[3] This value was corroborated in experiments by the same group where they observed vWF shape transition from a collapsed conformation to a stretched conformation above the same shear level.[53] Lippok *et al.* reported a similar value for the critical shear at which vWF becomes susceptible to cleavage by ADAMTS-13, indicating the onset of the conformational change.[35] While these studies examined vWF in simple shear flows with no streamwise acceleration/deceleration, it has been established that the velocity gradients in the direction of flow play a major role in vWF unfolding. In polymer dynamics simulations by Sing *et al.*, the unfolding threshold for a purely elongational flow field was two orders of magnitude lower than that for a simple shear.[60] In experimental studies by Nesbitt, Tovar-Lopez, Westein and colleagues, vWF-mediated platelet aggregation occurred only in the presence of strong flow gradients produced by a stenotic geometry.[40,69,71]

The mechanism of flow-induced polymer chain unfolding in dilute solutions of flexible polymers has been studied extensively in the context of drag-reducing additives.[38] In 1974, de Gennes proposed a phase diagram for predicting the state of dilute flexible polymers in general incompressible flows.[21] The velocity gradient tensor, $grad\ \boldsymbol{v}$, can be decomposed in a unique manner into its symmetric part, $\boldsymbol{D}$, and an anti-symmetric part, $\boldsymbol{\Omega}$, so that $grad\ \boldsymbol{v} = \boldsymbol{D} + \boldsymbol{\Omega}$, where $\boldsymbol{D} = \frac{1}{2}\big((grad\ \boldsymbol{v}) + (grad\ \boldsymbol{v})^T\big)$ and $\boldsymbol{\Omega} = \frac{1}{2}\big((grad\ \boldsymbol{v}) - (grad\ \boldsymbol{v})^T\big)$. $\boldsymbol{D}$ is often referred to as stretching tensor and $\boldsymbol{\Omega}$ is the spin tensor associated with rotation.[8,11,66] Their magnitudes are given by

$$\|\boldsymbol{D}\| = \sqrt{\frac{1}{2}\boldsymbol{D}:\boldsymbol{D}} \qquad \|\boldsymbol{\Omega}\| = \sqrt{\frac{1}{2}\boldsymbol{\Omega}:\boldsymbol{\Omega}} \qquad (1), (2)$$

de Gennes (1974) predicted that in arbitrary two-dimensional flows polymers subjected to high velocity gradients would exhibit a sharp coiled-to-stretched transition if $\|\boldsymbol{D}\| > \|\boldsymbol{\Omega}\|$ and remain coiled if $\|\boldsymbol{D}\| < \|\boldsymbol{\Omega}\|$, whereas in simple shear, where $\|\boldsymbol{D}\| = \|\boldsymbol{\Omega}\|$, the sharp transition would vanish and the polymer would be on the verge of instability.[4,21]

Many of the predictions made by de Gennes have since been confirmed in studies of DNA polymers in flow and Brownian dynamics simulations of flexible polymer chains.[3,26,27,33,45,54,55,63,72] In extensional flows such as planar or uniaxial extension, polymers undergo an abrupt steady-state unfolding above a critical strain rate $\dot{\epsilon}$. In simple shear, however, polymers undergo a tumbling motion where they extend and collapse periodically above a critical shear rate $\dot{\gamma}$. (The velocity gradients corresponding to the flows discussed above are provided in the SI Appendix 1.)

Since the regulation of the vWF action takes place through its conformational change, it is essential to account for the state of these polymer chains in any mathematical model of coagulation/thrombosis in high shear. We introduce herein a continuum model of vWF unfolding, which is developed within the framework of our multi-constituent model of platelet-mediated thrombosis, reported previously.[74,77] The vWF constituent is presented independently in this work and will subsequently be incorporated in the model of thrombosis to account for the vWF-mediated platelet deposition and aggregation.

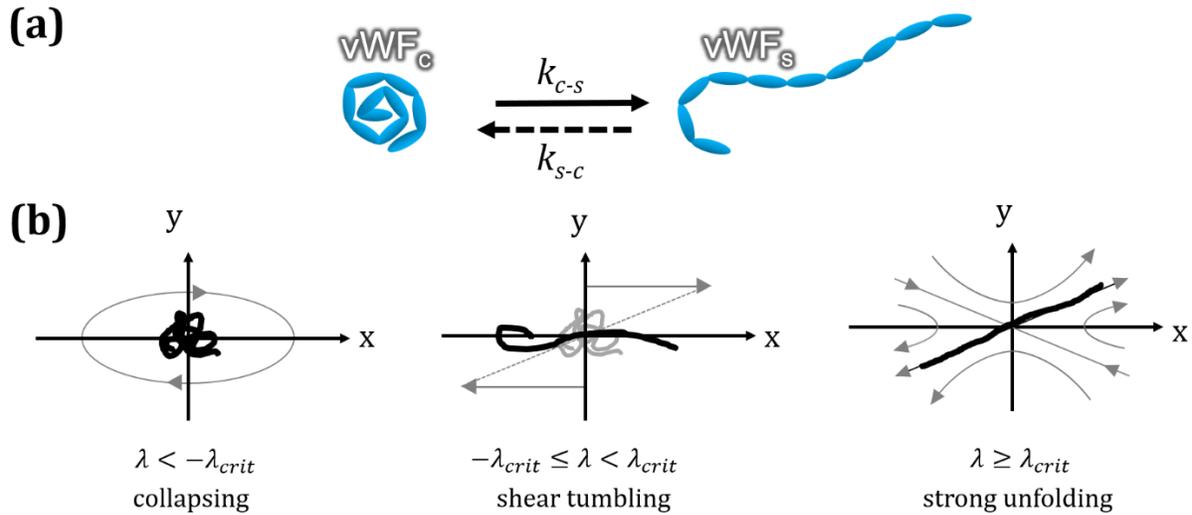

**Fig. 1.** (a) Schematic depiction of the two vWF species where $k_{c\text{-}s}$ and $k_{s\text{-}c}$ are the collapsed-to-stretched and stretched-to-collapsed conversion rates, respectively. (b) Polymer chain configuration in flows with various flowtype, $\lambda$ = [-1, 1], where $\lambda_{crit}$ = 0.0048.[4] When $\lambda$ < -$\lambda_{crit}$, the chain is collapsed. In near-shear flows, $\lambda \approx 0$, the chain undergoes a tumbling motion (unfolding-collapsing cycles). When $\lambda \geq \lambda_{crit}$, the chain is fully extended with minimal chain fluctuations. Adapted from Woo and Shaqfeh,[72] with the permission of AIP Publishing.

## Materials and Methods

### Equations of motion for blood

Blood is treated as a Newtonian incompressible fluid governed by the equations of conservation of mass and linear momentum:

$$div\ \boldsymbol{v} = 0 \tag{3}$$

$$\frac{D\boldsymbol{v}}{Dt} = \frac{1}{\rho} div\ \boldsymbol{T} \tag{4}$$

where $\rho$ is the blood density (1050 kg/m$^3$), $\boldsymbol{v}$ is the velocity; $\boldsymbol{T}$ is the Cauchy stress tensor represented by $\boldsymbol{T} = -p\boldsymbol{I} + 2\mu\boldsymbol{D}$, where $p$ is the pressure, $\boldsymbol{I}$ is the identity tensor, $\mu$ is the asymptotic viscosity (3.5 cP) of blood, and $\boldsymbol{D}$ is the symmetric part of the velocity gradient tensor.

### vWF transport and unfolding mechanism

We introduce two scalar species corresponding to the two conformational states of vWF polymers: collapsed polymers, [vWF$_c$], and stretched polymers, [vWF$_s$], depicted in Fig. 1a. The two states are interconvertible with conversion rates of $k_{c\text{-}s}$ (collapsed-to-stretched) and $k_{s\text{-}c}$ (stretched-to-collapsed) such that the total vWF concentration is conserved. (See Eqs. (5)-(6).) In all simulations, only the collapsed vWF is present initially, [vWF$_c$] = vWF$_{total}$, and [vWF$_s$] is produced by means of [vWF$_c$] unfolding. The transport and interconversion of the species in the flow field is described by a set of convection-diffusion-reaction equations:

$$\frac{d}{dt}[vWF_c] + div(\boldsymbol{v} \cdot [vWF_c]) = div(D_{vWF} \cdot \nabla[vWF_c]) - k_{c\text{-}s}[vWF_c] + k_{s\text{-}c}[vWF_s] \tag{5}$$

$$[vWF_c] + [vWF_s] = vWF_{total} \tag{6}$$

where $\boldsymbol{v}$ is the fluid velocity and $D_{vWF}$ = 3.19×10$^{-11}$ m$^2$s$^{-1}$ is the diffusivity[34] of vWF species.

### Flow classification and vWF behavior

Three polymer behavior regimes, illustrated in Fig. 1b, are defined based on the flow classification from Babcock *et al.* who studied the conformational response of polymers near the critical point of the coil-stretch transition.[4] The flowtype parameter, $\lambda$, is given by:

$$\lambda = \frac{\|\boldsymbol{D}\| - \|\boldsymbol{\Omega}\|}{\|\boldsymbol{D}\| + \|\boldsymbol{\Omega}\|} \tag{7}$$

Such that $\lambda$ = 1 in purely extensional flow, $\lambda$ = -1 in pure rotation, and $\lambda$ = 0 in simple shear. Babcock et al. found that in extension-dominated flows with flowtype values between 0 and 1, the average stretch of the polymer is determined solely by the largest eigenvalue of $grad\ \boldsymbol{v}$ since the polymer chain orients along the eigenvector corresponding to the extension axis. Therefore, they modified the Weissenberg number, $Wi = \tau_{rel}(\|\boldsymbol{D}\| + \|\boldsymbol{\Omega}\|)$ using the flowtype, $\lambda$, so that $\sqrt{\lambda}(\|\boldsymbol{D}\| + \|\boldsymbol{\Omega}\|)$ gives the largest eigenvalue of the velocity gradient:[16,20,41]

$$Wi_{\text{eff}} = \sqrt{\lambda} \cdot Wi = \tau_{rel}\sqrt{\lambda}(\|\boldsymbol{D}\| + \|\boldsymbol{\Omega}\|) \quad (8)$$

where $\tau_{rel}$ is the polymer relaxation time.

In simple shear, $\lambda = 0$, and hence $Wi_{\text{eff}} = 0$. In this regime, the configuration of a polymer never reaches a stable equilibrium at high $Wi$ since the extension and compression eigenvectors are parallel to each other, so the chain undergoes periodic extension and collapsing.[4] Similarly, in near-shear flows, $-\lambda_{\text{crit}} \leq \lambda < \lambda_{\text{crit}}$, the angle between the two axes is small enough for the Brownian motion of the monomers to induce tumbling.[56,72]

Thus, vWF will experience strong unfolding at high $Wi$ in flows with $\lambda \geq \lambda_{\text{crit}}$, shear tumbling in $-\lambda_{\text{crit}} \leq \lambda < \lambda_{\text{crit}}$, and remain collapsed if $\lambda < -\lambda_{\text{crit}}$. These three regimes are modeled by the conversion rates, $k_{c-s}$ and $k_{s-c}$, as described below. The critical value for flowtype, $\lambda_{\text{crit}} = 0.0048$, is adopted from the experiment of Babcock et al.[4]

*Shear tumbling*

In simple shear and near-shear flows, $-\lambda_{\text{crit}} \leq \lambda < \lambda_{\text{crit}}$, the unfolding rate, $k_{c-s}$ (collapsed-to-stretched conversion), follows a function proposed by Lippok et al. for shear-dependent cleavage rate of vWF [35]. Since this function describes the availability of vWF monomers for enzymatic cleavage the unfolding rate is assumed to follow the same shape, given by Eq. (9):

$$k_{c-s}^{\text{shear}}(\dot{\gamma}) = \frac{k'}{1 + \exp\left(-\frac{\dot{\gamma} - \dot{\gamma}_{1/2}}{\Delta \dot{\gamma}}\right)} \quad (9)$$

$$k_{s-c}^{\text{shear}} = k' \quad (10)$$

where $\dot{\gamma}_{1/2} = 5522$ s⁻¹ is the half-maximum shear rate for unfolding and $\Delta \dot{\gamma} = 1271$ s⁻¹ is the width of the transition from Lippok et al.[35]; $k'$ is the nominal state conversion rate, $k' = \frac{1}{t_{vWF}}$, where $t_{vWF} = 50$ ms is the vWF unfolding time. This timescale for vWF unfolding was reported in experiments by Fu et al.[19] and Brownian dynamics simulations by Dong et al.[15,28] Fig. 2 shows the plot of $k_{c-s}^{\text{shear}}$ normalized by $k'$ as a function of shear rate, $\dot{\gamma}$.

To reflect the tumbling behavior of the vWF in this regime, the stretched-to-collapsed conversion rate, $k_{s-c}$ is set to the nominal value, $k_{s-c}^{\text{shear}} = k'$, such that at high shear rates, when all vWF chains are undergoing shear tumbling, at any given moment of time, roughly half of the total vWF concentration would be in a stretched conformation and the other half in a collapsed conformation. This mirrors the experimental and simulation results where the ensemble-averaged fractional extension of vWF approached 0.5 with increasing shear rate.[3,37,53]

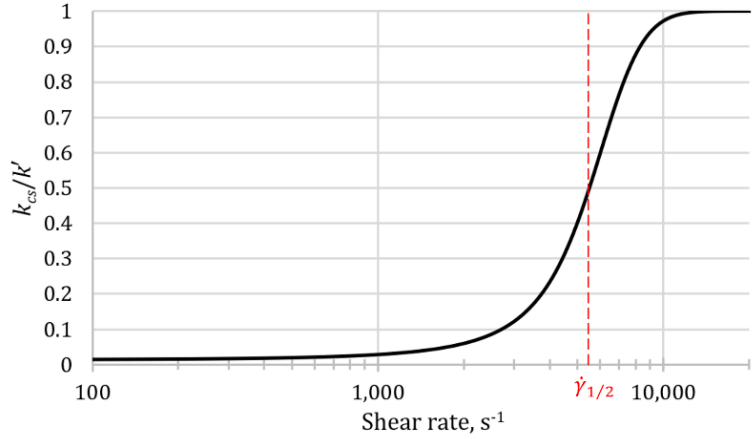

**Fig. 2.** Plot of the unfolding rate in simple shear, $k_{c\text{-}s}^{\text{shear}}$, normalized by the nominal state conversion rate, $k'$, as a function of shear rate, $\dot{\gamma}$. The half-maximum shear rate $\dot{\gamma}_{1/2}$ = 5522 s$^{-1}$ corresponds to the value around which vWF is commonly reported to unfold in simple shear.[3,35,53]

*Strong unfolding*

In extensionally-dominated flows with $\lambda \geq \lambda_{\text{crit}}$, strong unfolding occurs if $Wi_{\text{eff}}$, given by Eq. (8), exceeds the critical value $Wi_{\text{eff,crit}}$. Then, the unfolding rate $k_{c\text{-}s}^{\text{strong}}$ scales with $Wi_{\text{eff}}$ according to Eq. (11), while $k_{s\text{-}c}^{\text{strong}}$ is zero.

If $Wi_{\text{eff}}$ falls below $Wi_{\text{eff,crit}}$, the stretched vWF$_s$ will still experience hysteresis [54,62] and will not collapse back until $Wi_{\text{eff}} < Wi_{\text{eff,hyst}}$, as shown in Eq. (12).

$$k_{c\text{-}s}^{\text{strong}} = \begin{cases} k' \dfrac{Wi_{\text{eff}}}{Wi_{\text{eff,crit}}}, & \text{if } Wi_{\text{eff}} \geq Wi_{\text{eff,crit}} \\ k_{c\text{-}s}^{\text{shear}}, & \text{if } Wi_{\text{eff}} < Wi_{\text{eff,crit}} \end{cases} \quad (11)$$

$$k_{s\text{-}c}^{\text{strong}} = \begin{cases} 0, & \text{if } Wi_{\text{eff,hyst}} \leq Wi_{\text{eff}} < Wi_{\text{eff,crit}} \\ k', & \text{if } Wi_{\text{eff}} < Wi_{\text{eff,hyst}} \end{cases} \quad (12)$$

Here, we adopted the unfolding threshold and the hysteresis value reported by Sing and Alexander-Katz [62]. Assuming the monomer diffusion time of 1.02×10$^{-3}$ s for $\tau_{rel}$ in Eq. (8), $Wi_{\text{eff,crit}}$ = 0.316 and $Wi_{\text{eff,hyst}}$ = 0.053. Detailed explanation of the monomer diffusion time is provided in SI Appendix 2.

*Collapsed conformation*

If $\lambda < -\lambda_{\text{crit}}$, vWF$_c$ remains collapsed, $k_{c\text{-}s}$ = 0, and vWF$_s$ reverts to a globular state, $k_{s\text{-}c} = k'$.

The mathematical model was numerically implemented in an open-source finite volume software library OpenFOAM (The OpenFOAM Foundation Ltd). The simulations were performed using OpenFOAM v6 and post-processing was done in ParaView 5.6.

## Results

### vWF unfolding in Couette flow

To demonstrate the shear tumbling behavior of vWF, a planar Couette flow was modeled in a rectangular domain with a moving top wall and a stationary bottom wall. Periodic boundary conditions (BC) were applied at the inlet and outlet of the domain and no-slip BC was prescribed at the walls. The domain height in the *y*-direction was 0.2 mm, and the moving wall velocity was adjusted to produce shear rates of 3000, 5000, and 10,000 s$^{-1}$. Timestep size was 2.5×10$^{-6}$ s. After obtaining the steady-state flow solution, the domain was initialized with [vWF$_c$] = vWF$_{total}$ = 1000 nmol/m$^3$ and [vWF$_s$] = 0.

Since $\lambda$ = 0 in simple shear flow, vWF unfolding takes place via shear tumbling where $k_{S-C}^{shear} = k'$ and $k_{C-S}^{shear}$ is a function of shear rate as per Eq. (9). The time evolution of the two conformational states of vWF at each shear rate is shown in Fig. 3. The steady-state concentration of vWF$_s$ normalized by the total vWF concentration, [vWF$_s$]/vWF$_{total}$, was roughly 0.1 at 3000 s$^{-1}$, 0.28 at 5000 s$^{-1}$, and 0.49 at 10,000 s$^{-1}$. With further increase in shear rate, the steady-state value of [vWF$_s$]/vWF$_{total}$ approaches 0.5 asymptotically. Fig. 4 shows the percentage of stretched vWF, ([vWF$_s$]/vWF$_{total}$)×100%, as a function of shear rate and exposure time.

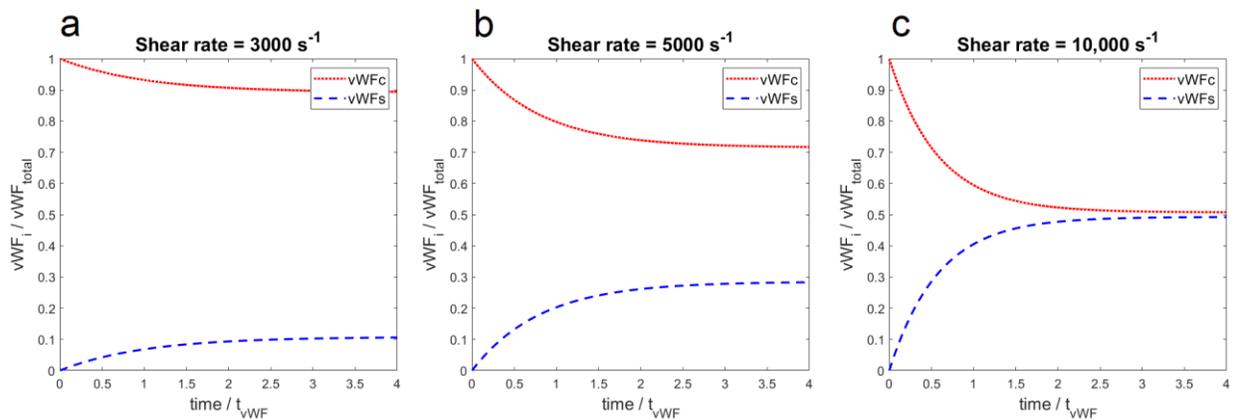

**Fig. 3.** Time evolution of the two conformational states, collapsed vWF$_c$ and stretched vWF$_s$, in simple shear at shear rates of 3000 (a), 5000 (b), and 10,000 s$^{-1}$ (c). vWF concentrations are normalized by vWF$_{total}$, and the time axis is normalized by $t_{vWF}$. At the greatest shear level, roughly half of the vWF is collapsed and half is stretched, mirroring the shear tumbling behavior observed experimentally, where the mean fractional extension of polymers approaches 0.5 with increasing shear rate.[53]

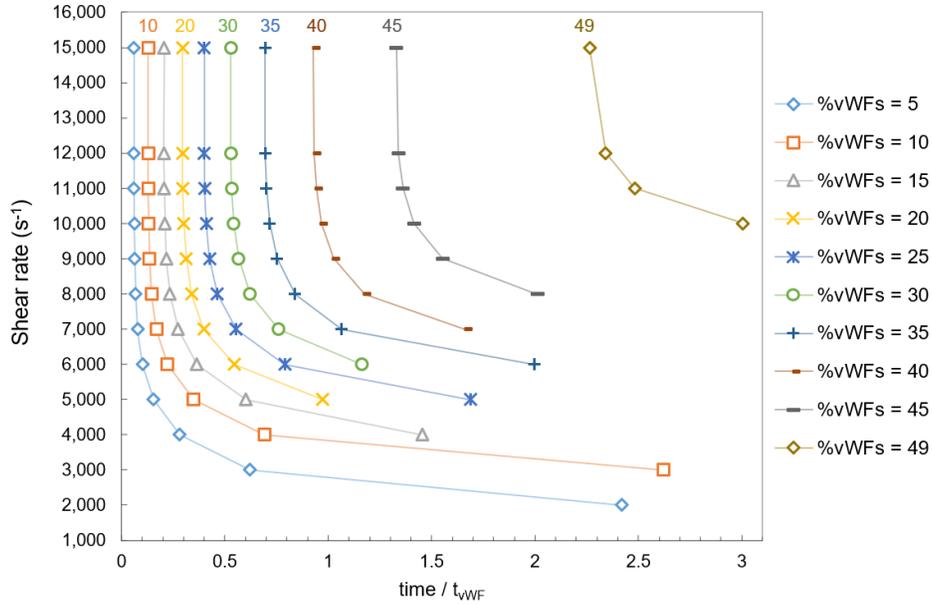

**Fig. 4.** Unfolding of vWF subjected to simple shear. Percentage of stretched vWF, ([vWF$_s$]/vWF$_{total}$)×100%, shown as a function of shear rate and exposure time. Time axis is normalized by $t_{vwf}$.

## vWF unfolding in a cross-slot channel

The flow within a 2-D cross-slot geometry was simulated to produce a planar extensional flow field.[23] The flow domain consists of mutually bisecting orthogonal rectangular channels of width $w$ = 0.4 mm, as shown in Fig. 5a. The geometry has two diametrically opposed inlets at the top and bottom ($y$-direction) and two opposing outlets exiting left and right ($x$-direction). A fully developed velocity profile was prescribed at the two inlets with the mean velocity of 0.2 m/s. Resulting velocity streamlines are hyperbolic, forming a stagnation point at the exact center of the intersection. After obtaining the steady-state flow solution, the domain was initialized with [vWF$_c$] = vWF$_{total}$ = 1000 nmol/m$^3$ and [vWF$_s$] = 0, and the same values were prescribed as Dirichlet BCs at the inlets.

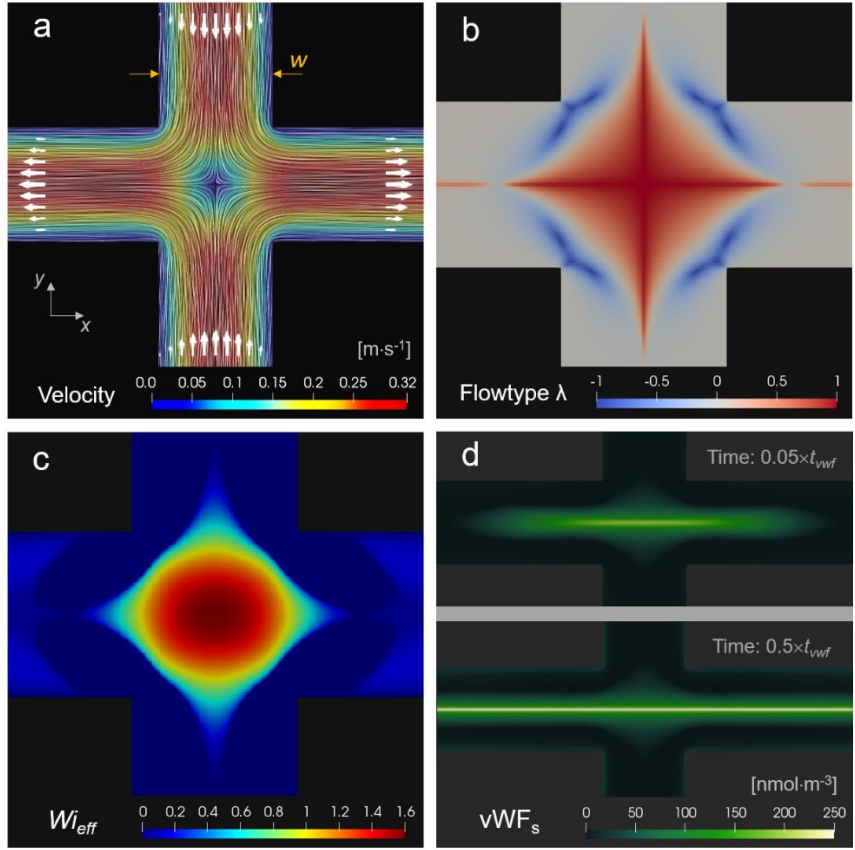

**Fig. 5.** (a) Velocity line integral convolution colored by velocity magnitude. Flow direction is indicated by velocity vectors. Channel width, *w* = 0.4 mm. (b) Flowtype parameter, λ, where λ = 1 corresponds to purely extensional kinematics, 0 is simple shear, and -1 is pure rotation. (c) Plot of $Wi_{eff}$ shows the greatest value at the central stagnation region. (d) Stretched $vWF_s$ concentration peaks along the outflow axis and resembles a typical polymer birefringent strand produced in cross-slot extensional rheometers.

The flow field approximates ideal planar extension in the region surrounding the central stagnation point and along the orthogonal symmetry axes, as demonstrated by the flowtype parameter, λ, in Fig. 5b. Indeed, the deformation rates at the central stagnation point are $\partial v_x/dx = 1567$ s$^{-1}$ and $\partial v_y/dy = -1567$ s$^{-1}$. $Wi_{eff}$ is also greatest at that location, $Wi_{eff}$ = 1.596 (Fig. 5c). From Eq. (8), dividing the $Wi_{eff}$ by $\tau_{rel}$ recovers the largest eigenvalue of the *grad(v)*. Using $\tau_{rel}$ = 1.02×10$^{-3}$ s yields $\sqrt{\lambda}(\|\mathbf{D}\| + \|\mathbf{\Omega}\|) \approx 1565$ s$^{-1}$.

In the regions where $Wi_{eff}$ exceeds the critical value $Wi_{eff,crit}$, the vWF chains undergo strong unfolding. The resulting concentration of the stretched $vWF_s$ is shown in in Fig. 5d. The concentration peaks along the outflow axis and resembles a typical polymer birefringent strand produced in cross-slot extensional rheometers [23]. Lower concentration of $vWF_s$ is present along the channel walls. In the near-wall regions, the flow approximates simple shear, λ ≈ 0, and vWF unfolding takes place via the shear tumbling regime.

## vWF unfolding in stenotic microfluidic systems

To verify the vWF unfolding model, flow was simulated in three microfluidic systems designed for inducing high-shear vWF-mediated thrombosis and screening for von Willebrand Disease.

### *Stenotic tube*

Ku and colleagues[5,44] studied thrombosis under arterial flow conditions employing a tube with a stenotic test section that mimics the shape of an atherosclerotic plaque, creating hemodynamic conditions of large shear rate gradients, extensional flow, high shear rates, and short exposure times. This experimental setup produced thrombotic occlusion at the throat of the stenosis within 10-15 minutes of whole blood perfusion.

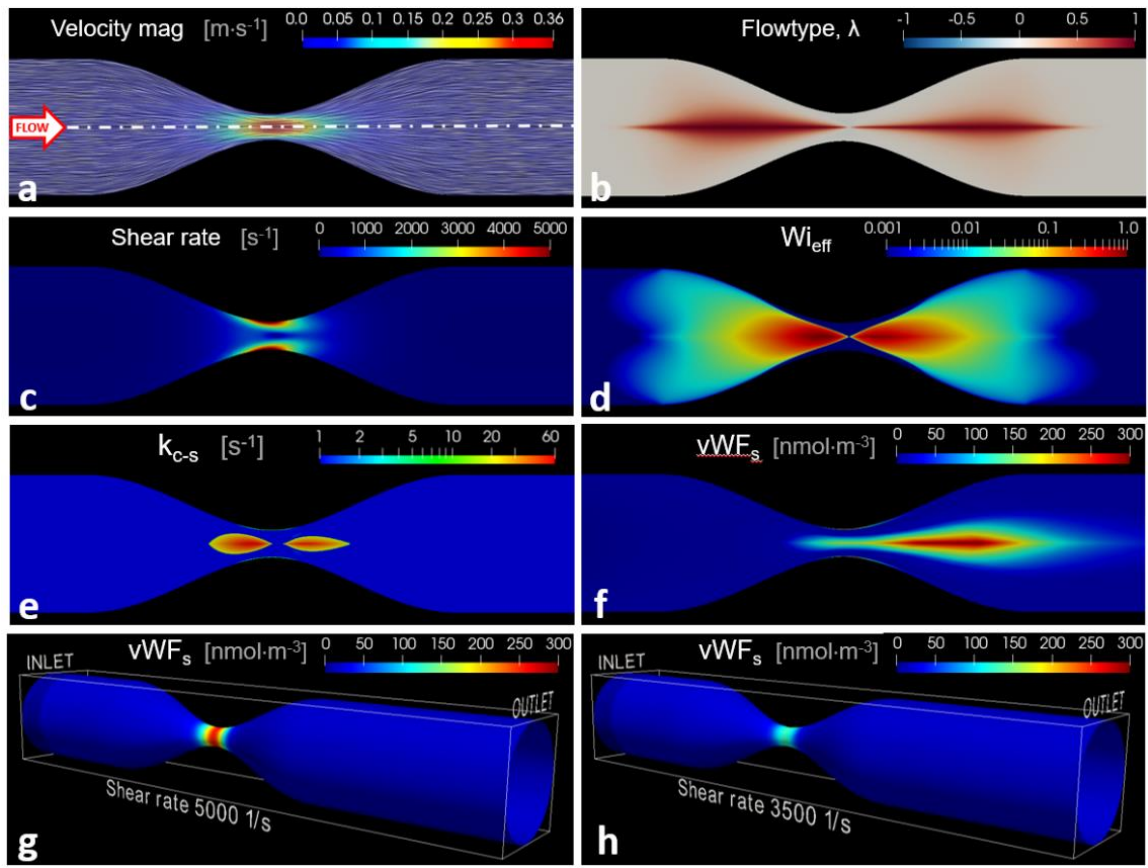

**Fig. 6.** Flow in a stenotic tube with 80% reduction in diameter. (a) Velocity line integral convolution colored by velocity magnitude. (b) Flowtype parameter, λ, reveals strong extensional kinematics along the center axis of the tube. (c) Peak shear rate at the throat of the stenosis reaches 5000 $s^{-1}$. (d)-(e) The magnitude of the $Wi_{eff}$ dictates the strong vWF unfolding with the collapsed-to-stretched conversion rate of $k_{c-s}$. (f) High $vWF_s$ concentration produced by strong unfolding near the tube center axis. (g)-(h) High $vWF_s$ concentration shown on the tube wall, produced by shear tumbling in the near-wall regions. Two flow rates were simulated corresponding to peak wall shear rates of 5000 and 3500 $s^{-1}$ at the throat of the stenosis.

Flow in an axisymmetric tube with 80% stenosis was simulated, where the initial diameter was 1.5 mm and the diameter at the throat of the stenosis was 0.3 mm. Following the published experimental conditions, the flow rate was adjusted to generate shear rates of 5000 [5] and 3500 [44] s$^{-1}$ at the narrowest section of the tube. After obtaining the steady-state flow solution, the domain was initialized with [vWF$_c$] = vWF$_{total}$ = 1000 nmol/m$^3$ and [vWF$_s$] = 0, and the same values were prescribed as Dirichlet BCs at the inlets.

Fig. 6a-f show the velocity and flow parameters on a middle cross section of the tube for the 5000 s$^{-1}$ case. The contracting and expanding sections of the stenosis create strong extensional kinematics along the centerline of the tube, demonstrated by the λ (Fig. 6b) and the $Wi_{eff}$ (Fig. 6d). The strong unfolding produces a high concentration of vWF$_s$ extending downstream of the stenosis (Fig. 6f). The elevated wall shear rate at the throat of the stenosis (Fig. 6c) induces vWF unfolding via shear tumbling. Fig. 6g and Fig. 6h show the vWF$_s$ concentration on the tube wall at shear rates of 5000 s$^{-1}$ and 3500 s$^{-1}$, respectively. The regions of high vWF$_s$ concentration on the wall correspond to the section of the tube where occlusive thrombi formed in the experiments of Ku and colleagues.[5,44]

*PFA-100*

The platelet function analyzer PFA-100® (Siemens Erlangen, Germany) is an in vitro system used as a screening test for patients with impaired primary hemostasis.[59] PFA -100® is highly sensitive to the influence of plasma vWF and is routinely used for screening for von Willebrand disease.[18,22,32] Fig. 7a shows a cross-section sketch of the PFA-100® test cartridge whose main component is a bio-active membrane with a central orifice of 140 μm diameter. During the test, whole blood is aspirated through a capillary towards the membrane (bottom to top), and as blood flows through the central orifice of the membrane, a thrombus forms until occlusion is achieved.

Flow was simulated in an axisymmetric domain mimicking the PFA-100® test cartridge. A developed velocity profile with the mean velocity of 0.0637 m/s was prescribed at the inlet to achieve the peak shear rate of 6000 s$^{-1}$ in the membrane orifice.[32] The initial and boundary conditions for the vWF species were identical to the previous case.

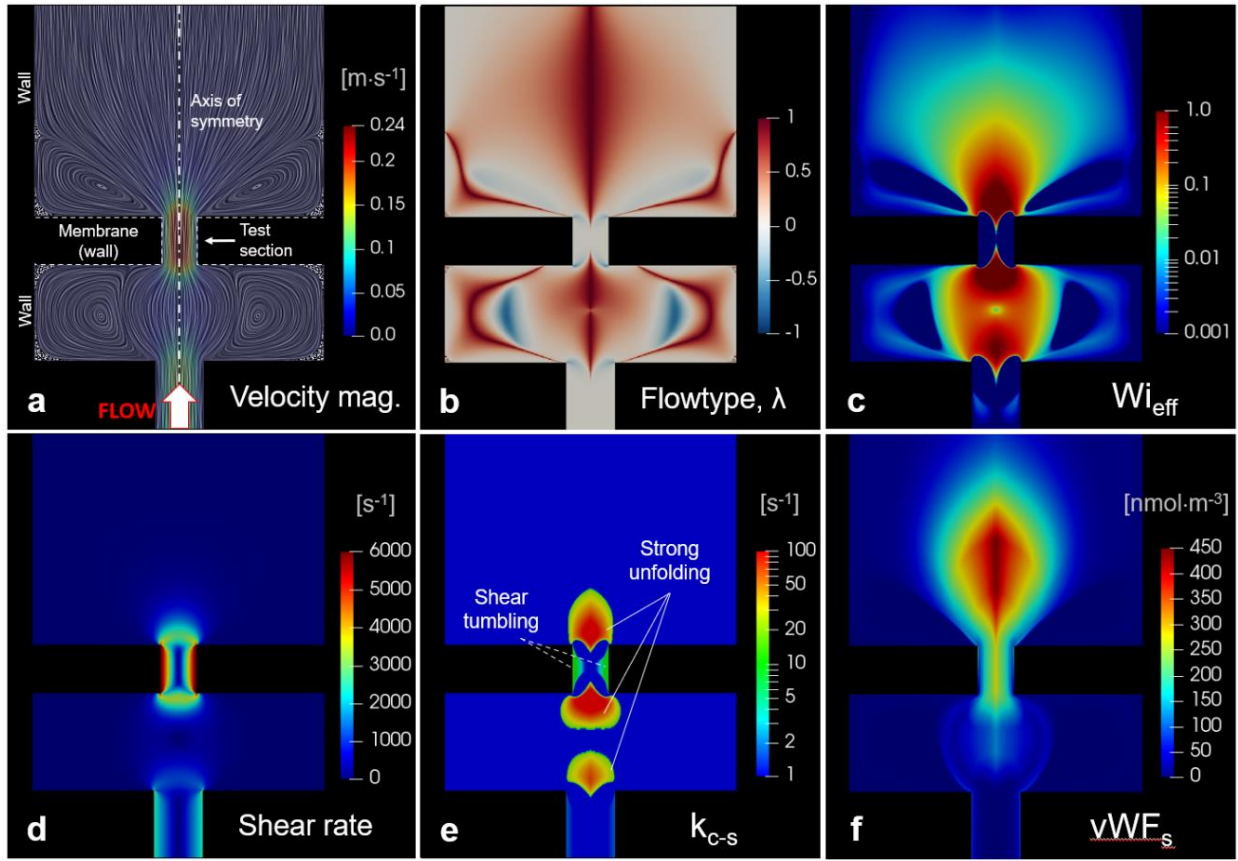

**Fig. 7.** Flow in the PFA-100® blood screening system. (a) Velocity line integral convolution colored by velocity magnitude. Flow direction is bottom to top. (b) Flowtype λ and (c) $Wi_{eff}$ reveal strong extensional kinematics in the expansion and contraction sections of the chambers. (d) Wall shear rate within the membrane orifice reaches 6000 s$^{-1}$. (e) The vWF unfolding rate $k_{c\text{-}s}$ shows three distinct regions of strong unfolding before and after the membrane as well as near-wall shear tumbling within the orifice. (f) High vWF$_s$ concentration is present in the region where vWF-mediated thrombotic occlusion is expected to occur.

The strong flow gradients created when the flow exits the capillary and is forced through the test orifice produces three distinct regions of strong vWF unfolding: at the exit of the capillary and the leading and trailing edges of the orifice. High wall shear rate inside the test section induces shear tumbling (Fig. 7e). The resulting concentration of the stretched vWF$_s$ is shown in Fig. 7f, which is critical for the process of thrombotic occlusion of the membrane orifice.

### Stenotic microfluidic channel

Westein *et al.* developed microfluidic chambers mimicking the geometry features and flow conditions occurring in stenosed carotid arteries in humans.[71] The flow chamber, shown in Fig. 8e, featured a half-circular eccentric stenosis with a diameter of 600 μm causing an 80% lumen reduction. Perfusion of whole human blood through this channel resulted in formation of large platelet aggregates in the outlet (expanding) region of the stenosis in a vWF-dependent manner.

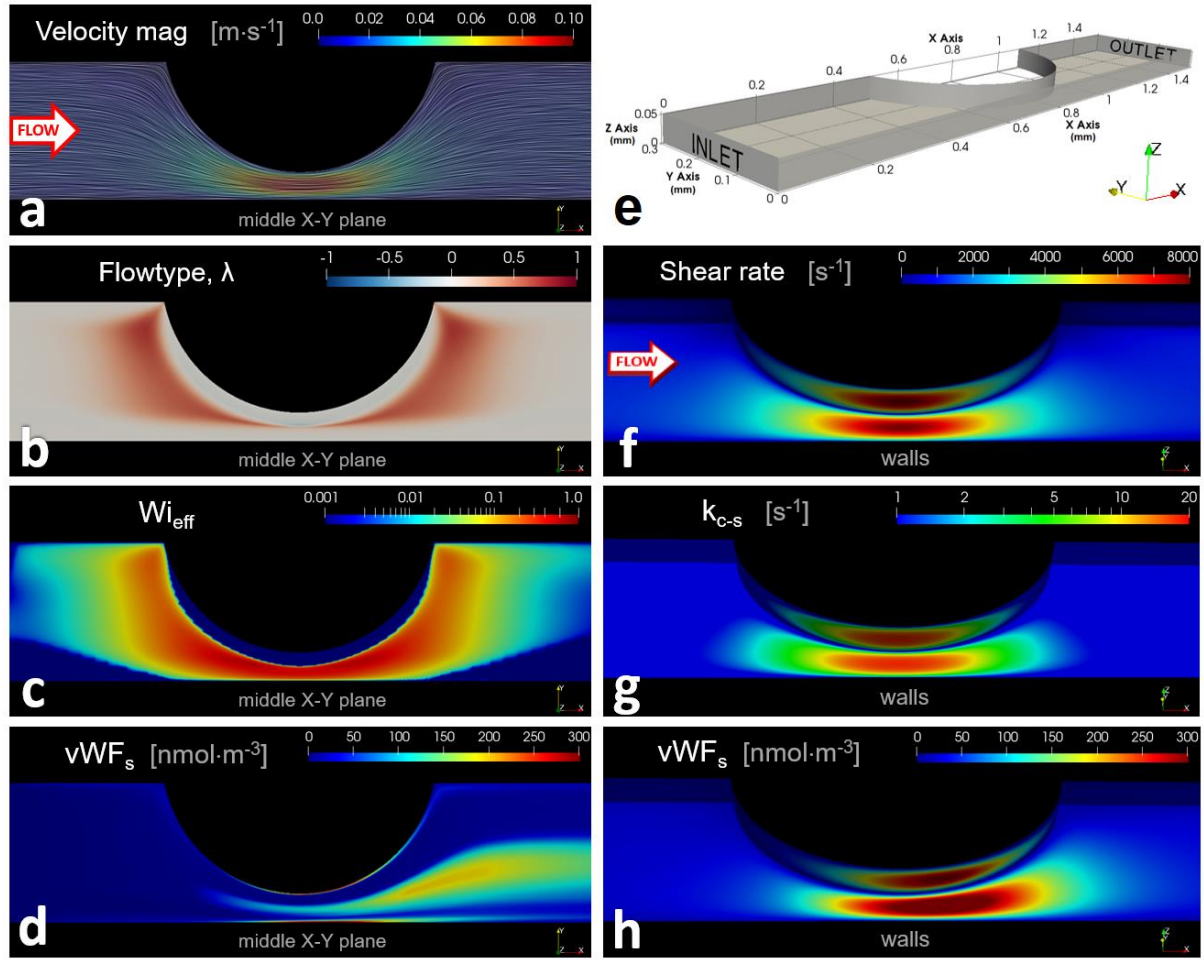

**Fig. 8.** Flow in a microfluidic channel with 80% stenosis. (a) Velocity line integral convolution shown on the middle X-Y plane, colored by velocity magnitude. (b) Flowtype λ and (c) $Wi_{eff}$ reveal strong extensional kinematics throughout the stenotic region. (d) Unfolded $vWF_s$ produced by strong unfolding near the middle X-Y plane of the channel. (e) 3-D schematic of the simulated flow chamber. (f) Wall shear rate and (g) unfolding rate due to shear tumbling, shown on the bottom and side walls of the channel. (h) High $vWF_s$ concentration along the channel walls extends from the apex of the stenosis toward the outlet.

The flow simulation mirrored the experimental conditions where the inlet flow rate was set at 0.5 ml/hr, producing an input wall shear rate of 1000 $s^{-1}$ that increased to 8000 $s^{-1}$ in the apex of the stenosis (Fig. 8f). Large streamwise velocity gradients generated by acceleration/deceleration of the flow through the stenotic section induce vWF unfolding near the middle X-Y plane of the channel (Fig. 8a-d). Near the walls, however, vWF unfolding occurs due to the high shear rates (Fig. 8f-g), producing a trail of peak $vWF_s$ concentration extending from the apex of the stenosis toward the outlet (Fig. 8h). Likewise, the experiments of Westein *et al*. exhibited enhanced platelet aggregation in the region downstream of the apex of the stenosis.

## Discussion

Although there is a great body of knowledge produced on the dynamics of loosely coiled polymers in good-solvent or theta-solvent conditions,[64] one must consider the critical difference of vWF from those polymers. Strong attractive interactions between its monomers give vWF a robust globular conformation, imparting significantly higher and well-defined thresholds for unfolding.[53,61] For this reason, although Babcock *et al.* observed modest deformation of DNA molecules at negative values of the flowtype $\lambda$,[4] we assume that the strong collapsed conformation of vWF would prevent any size fluctuation in this regime. However, once unfolding is initiated, vWF exhibits the same two types of behavior: unfolding-collapsing cycles in simple shear and abrupt but stable unfolding in extensional flow. These have been observed in single-molecule imaging experiments as well as hydrodynamic simulations of vWF.[53,60–62,65] Therefore, we adopted the flow classification framework from mixed flow studies of coiled polymer chains,[4,25,26,57,72] which aligns with the vWF nucleation- protrusion theory developed by Sing and Alexander-Katz,[62] and we used the unfolding thresholds from vWF literature.[3,35,53,60,62]

The shear tumbling behavior in the model is achieved through simultaneous action of the stretched-to-collapsed and collapsed-to-stretched conversion rates, $k_{s-c}$ and $k_{c-s}$, respectively. In this regime, $k_{s-c}$ is constant, meaning the default behavior of vWF chains is to collapse into a globular state. $k_{c-s}$, on the other hand, is a function of shear rate, with its asymptotic value matching $k_{s-c}$ (Fig. 2). Thus, at high shear rates, this results in an equal steady-state distribution of vWF$_c$ and vWF$_s$ concentrations (Fig. 3c), reflecting the fact that when all vWF chains are undergoing shear tumbling, at any given moment of time roughly half of the total vWF concentration would be in a stretched conformation and the other half in a collapsed conformation. This result also mirrors the steady-state mean fractional extension of 0.5 observed experimentally with polymers in simple shear.[53,63] Unfolding of a portion of the total vWF at sub-critical shear rates (Fig. 3a) is explained by the exponential size distribution of the vWF multimers and the higher susceptibility of the longer chains for unfolding.[34,42,76]

When subjected to flow with a dominant extensional component, a polymer chain aligns with the principal axis of extension and experiences steady unfolding, as shown in the right frame of Fig. 1b. Therefore, the collapsing rate $k_{s-c}$ is zero, and the unfolding rate $k_{c-s}$ scales with the $Wi_{eff}$, which is proportional to the eigenvalue of the velocity gradient along this principal axis.

In most flows in stenotic geometries both unfolding mechanisms will be present as the sudden changes in lumen area often create strong streamwise velocity gradients as well as elevated wall shear rates in the narrowed sections of the flow path. In the stenotic tube mimicking the hemodynamics of arterial atherosclerosis (Fig. 6), strong unfolding takes place along the central axis of the tube immediately upstream and downstream of the throat of the stenosis. However, the flow near the walls of the tube approximates simple shear, and therefore the initial platelet deposition at the walls would rely on the vWF unfolding via the shear tumbling mechanism. Fig. 6g and Fig. 6h show the high vWF$_s$ concentration in the region of the tube wall where occlusive thrombi formed in the experiments of Ku and colleagues.[5,44] Similarly, in the stenotic flow chamber of Westein *et al.* (Fig. 8), strong unfolding is limited to the middle plane of the channel, whereas shear tumbling is responsible for the trail of high vWF$_s$ concentration on the bottom wall, where enhanced platelet aggregation was observed in the experiment.[71] In the PFA-100® blood screening device, both unfolding mechanisms take place upstream of the membrane and within the orifice, resulting in a high concentration of stretched vWF within the test section (Fig. 7). It must be noted that vWF unfolding is a necessary but not a sufficient condition for high-shear thrombus formation. Multi-constituent thrombosis simulations need to be performed to enable a direct comparison to the experimental results and to demonstrate the role of vWF in supporting platelet aggregation under pathologic flow conditions.

Several simplifying assumptions and limitations of this work must be discussed. The binary conformational states of the vWF multimers in the model, collapsed or stretched, do not reflect partial unfolding that may be possible at intermediate flow strengths. The mechanism of vWF self-assembly leading to vWF net formation was not considered in this work.[31] The effect of surface-enhanced unfolding of collapsed polymers near walls shown by Alexander-Katz and Netz was omitted,[1] and the contribution of the vWF unfolding to the stress tensor of the fluid was ignored.[61] The unfolding and hysteresis thresholds in the strong unfolding regime, $Wi_{eff,crit}$ and $Wi_{eff,hyst}$, were simplified as constant values, whereas these thresholds would be lower for high-molecular-weight vWF concatemers. The vWF unfolding rate in simple shear (Eq.(9)) was inferred from the experiments on the enzymatic cleavage of vWF chains[35] and the nominal state conversion rate, $k'$, was obtained with the assumption of a typical vWF unfolding time, $t_{vWF}$ = 50 ms, as reported by Fu *et al*.[19] However, the latter experimental observations were made with wall-tethered vWF chains in a high-viscosity medium. Further experimental data are needed to better approximate the unfolding kinetics. Alternatively, these kinetic rates can be computed through Brownian dynamics[3] and Langevin dynamics[37] simulations in the future. Additionally, a constant value of $t_{vWF}$ is assumed in this work, whereas physically the unfolding time would be shear dependent. Ongoing work focuses on incorporating the present vWF model into our multi-constituent model of thrombosis[74,77] to account for the vWF-mediated platelet deposition and aggregation. To assess the sensitivity of the combined model to individual parameters, such as $t_{vWF}$, we are performing a global sensitivity analysis using Sobol indices computed with a polynomial chaos expansion as a parametric surrogate for the thrombosis model.

The macro-scale treatment of vWF adopted in this work was made possible by rigorous molecular-level simulation studies of vWF. Brownian dynamics (BD) simulations by Sing, Alexander-Katz, and Netz demonstrated the differing responses of vWF multimers to elongational and simple shear flows, as well as corresponding unfolding thresholds.[2,60,62] Liu *et al.* developed a multiscale computational model based on a coupled lattice-Boltzmann and Langevin-dynamics method where the suspension dynamics and interactions of individual platelets and VWF multimers are resolved directly.[36,37] It should be noted that vWF conformational change is the initial step to the subsequent processes: vWF binding to platelet GP1bα requires a mechanical force-activation of the vWF A1 domain, and vWF cleavage by ADAMTS13 requires further (local) unfolding of the A2 domain.[19,76] Dong *et al.* and developed a coarse-grained model that employed spring systems to mimic the mechanical force-extension response of the vWF A2 domain.[15] This model captured the polymer-level behavior (vWF conformational change) and the monomer-level behavior (A2 domain unfolding) at the same time.

While molecular-level simulations can provide valuable insight into polymer behavior, simulating the individual response of vWF multimers in physiologic concentrations flowing through complex flow environments such as full-scale medical devices would become prohibitively expensive, especially if combined with a numerical model of coagulation/thrombosis. The continuum approach employed in this work follows the framework of our multi-constituent model of thrombosis[74,77] which has demonstrated its feasibility in full-scale LVAD simulations.[75]

Caution is warranted when extending the flow classification method used here to arbitrary 3-D flows and turbulent flows. In 2-D flows, the vorticity vector is confined to a plane perpendicular to the velocity field and no vortex stretching occurs. However, the vortex stretching mechanism is essential to the 3-D structure of turbulence and the transfer of energy from large to small scales.[12,67] Therefore, alignment of vorticity and strain-rate eigenvectors is likely to cause stretching of polymer chains in the vorticity direction.[14,72] Moreover, in isochoric (incompressible fluid) flows, the sum of the eigenvalues of the velocity gradient vanishes according to the continuity equation, which means that a compression axis/plane is always associated with an extension plane/axis in 3-D flows. In other words, even if the real eigenvalue is negative (compression), the real part of the two other complex-conjugate eigenvalues is positive. For this reason, Terrapon *et al.* proposed assessing the positive real part of the eigenvalues of the local velocity gradient tensor as a measure of the ability of the flow to stretch the polymer.[68] Even then, since the local velocity gradient in a turbulent flow constantly changes with time, they hypothesized that only strong events such as biaxial extension are likely to cause polymer stretching.

For reasons discussed above, any computational model of vWF stretching and degradation in turbulent flows would also be limited by the resolution and accuracy of the flow solution. Bortot *et al.* 2019 showed that turbulence promotes vWF scission by ADAMTS13 and suggested that large intermittent transient forces occurring near the Kolmogorov scale are responsible for vWF degradation.[9] Sharifi and Bark estimated the forces acting on the vWF chain in turbulent flows and showed that the smallest scale turbulent structures have the biggest effect on tensile forces within vWF.[58] For cases in which it is infeasible to resolve fine scales of turbulent motion, development of surrogate models for vWF degradation could be a viable alternative, similar to those proposed for red blood cell damage in turbulent flows.[17,24,30,43,73]

## Acknowledgements

Author Wei-Tao Wu thanks the support of the grant NSFC 11802135. This work was supported by the National Institute of Health grant R01HL089456.

Authors are grateful to Dr. Mahdi Esmaily Moghadam (Sibley School of Mechanical and Aerospace Engineering, Cornell University) for valuable discussions on turbulent flows.

# Supplementary Information

## Appendix 1

The most commonly studied examples of extensional flows are planar extension and uniaxial extension.[1] Writing the transpose of the velocity gradient tensor, $\boldsymbol{\kappa}$, such that only the diagonal components are nonzero:

$$\boldsymbol{\kappa} = \begin{pmatrix} \dot{\epsilon} & 0 & 0 \\ 0 & -\dot{\epsilon} & 0 \\ 0 & 0 & 0 \end{pmatrix} \quad \text{(planar extension)}$$

$$\boldsymbol{\kappa} = \begin{pmatrix} \dot{\epsilon} & 0 & 0 \\ 0 & -\dot{\epsilon}/2 & 0 \\ 0 & 0 & -\dot{\epsilon}/2 \end{pmatrix} \quad \text{(uniaxial extension)}$$

The velocity gradient tensor for a simple shearing flow contains an asymmetric off-diagonal term, the shear rate $\dot{\gamma}$:

$$\boldsymbol{\kappa} = \begin{pmatrix} 0 & \dot{\gamma} & 0 \\ 0 & 0 & 0 \\ 0 & 0 & 0 \end{pmatrix} \quad \text{(simple shear)}$$

## Appendix 2

Sing and Alexander-Katz report the vWF unfolding thresholds from their simulations in a nondimensionalized form using *Wi*, where for extensional flow, $\dot{\epsilon}\tau_{mon} \approx 0.316$, and for simple shear, $\dot{\gamma}\tau_{mon} \approx 5.62$.[2] They use the monomer diffusion time, $\tau_{mon}$, which is calculated as:

$$\tau_{mon} = \frac{a^2}{\mu_0 k_B T}, \qquad \mu_0 = \frac{1}{6\pi\eta a}$$

where *a* is vWF monomer radius, $\eta$ is the solvent dynamic viscosity, $k_B$ is Boltzmann constant, and *T* is absolute temperature in K. Therefore, the value of $\tau_{mon}$ will depend on the assumption of the monomer radius, *a*. Sing and Alexander-Katz assumed *a* = 50 nm in their prior work.[3] The vWF repeating unit has been reported to be around 70 nm along its major axis.[4,5] In the interest of consistency, we used the simple shear unfolding threshold from Eq. (9), $\dot{\gamma}_{1/2}$ = 5522 s$^{-1}$, to compute $\tau_{mon}$ from $\dot{\gamma}\tau_{mon} \approx 5.62$. To verify the resulting monomer diffusion time of $\tau_{mon}$ = 1.02 ms, we compute the corresponding monomer radius *a*. Assuming blood plasma at 37°C, $\eta$ = 0.0012 kg·m$^{-1}$s$^{-1}$, T = 310.15 K, the monomer radius is *a* = 58 nm, which falls within the range of values mentioned above.